\documentstyle[12pt,epsf,moriond]{article}

\begin{document}
\newcommand{\magcir}{\ \raise -2.truept\hbox{\rlap{\hbox{$\sim$}}\raise5.truept
 	\hbox{$>$}\ }}		

\def\araa{{\it Ann. Rev. Astr. Ap.}\ }
\def\apj{{\it ApJ}\ }
\def\apjs{{\it ApJS}\ }
\def\aj{{\it AJ}\ }
\def\aaa{{\it A\&A}\ }
\def\aas{{\it A\&AS}\ }
\def\mnras{{\it MNRAS}\ }
\def\nat{{\it Nat}\ }

\parindent 15pt

\heading{THE SUNYAEV-ZELDOVICH EFFECT \\AT 1 AND 2 MM TOWARDS ROSAT CLUSTERS}

\author{P. Andreani$^1$, G. Dall'Oglio$^2$, L. Pizzo$^2$, L. Martinis$^3$,
P. Shaver$^4$, H. B\"ohringer$^5$, R. Lemke$^4$, L.-\AA.~ Nyman$^4$,
R. Booth$^6$, N. Whyborn$^7$}
{$^1$Dip.Astr.,Univ. Padova, Italy; $^2$Dip.Fis.,III Univ. Roma, Italy}
{$^3$ENEA Frascati, Italy; $^4$ESO Garching, Germany}
{$^5$MPE Garching, Germany; $^6$ONSALA Sweden; $^6$SRON, Groningen}

%\author{P. Andreani$^1$, G. Dall'Oglio$^2$, L. Pizzo$^2$, L. Martinis$^3$,
%P. Shaver$^4$, H. B\"ohringer$^5$, R. Lemke$^6$, L.-\AA.~ Nyman$^6$,
%R. Booth$^7$, N. Whyborn$^8$}
%{$^1$Dip. di Astronomia, Univ. di Padova, Italy\\
%$^2$Dip. di Fisica, III Univ. Roma, Italy\\
%$^3$ENEA TIB, Frascati, Italy\\
%$^4$ESO Garching, Germany\\
%$^5$MPE Garching, Germany\\$^6$ESO La Silla, Chile\\
%$^7$ONSALA Observatory, Sweden\\$^8$SRON, Groningen, The Netherlands}

\begin{abstract}{\baselineskip 0.4cm 
An observing campaign was devoted to the search for 
the Sunyaev-Zeldovich (S-Z) effect towards
X-ray ROSAT Clusters in the millimetric spectral domain.
A double channel (1.2 and 2 {\it mm}) photometer was
installed at the focus of the 15m Swedish ESO Submillimeter Telescope (SEST)
in Chile in september 1994 and 1995 and observations of the targets S1077,
A2744, S295 and RXJ0658-5557 were gathered.
Detections were found for A2744 at 1 {\it mm} and in both channels
(at 1.2 and 2 {\it mm}) towards RXJ0658-5557. For the first time there is
evidence for the S-Z enhancement and both the latter
and the decrement were detected on the same source.
We discuss astrophysical and systematic effects which could give origin to these
signals.}

\end{abstract}

\section{Introduction}
The S-Z effect is one of the major sources of 
secondary anisotropies of the Cosmic Microwave Background (CMB), arising from (inverse) Compton
scattering of the microwave photons by hot electrons in clusters of galaxies.
This effect generates a peculiar signal with a decrement at wavelengths longer than
1.4{\it mm} and an enhancement at shorter ones relative to the CMB
planckian value.

The original computation by Sunyaev and Zeldovich $^{[1-3]}$
of the the net transfer of energy from the hot e$^-$ to the microwave
photons predicts a signal for the
relative temperature change:
\begin{equation}
({\Delta T \over T})_{therm} = ~y~(x ~{{ e^x +1 }\over {e^x -1}} -4)
\label{eq:curv}.
\end{equation}

\noindent
where T is the CMB temperature, $x=h\nu/kT$ and $y=\int (kT_e/mc^2)~
n_e \sigma _T d\ell$ is the comptonization parameter, $n_e$, T$_e$ being
the electron density and temperature. Equation 1 is an approximated
solution of the full
kinetic equation for the change of the photon distribution due to scattering.
A more accurate solution gives rise to corrections which are not
negligible at high frequencies $^{[4]}$.

If the cluster has a
peculiar velocity relative to the frame where the CMB is isotropic
 an additional {\it kinematic} effect should be measured.
The motion of the gas cloud will induce a Doppler change whose
relative amplitude, $({\Delta T \over T} )_{kin}$, does not depend on the
frequency but only on the peculiar velocity, $v_r$, and cloud optical depth
for Thomson scattering, $\tau$,: $({\Delta T \over T})_{kin} = -
{v_r \over c} \tau $
(where the minus sign refers to a cluster receding from the observer).
Since both effects are very small the net relative temperature change
is the sum of the two: $({\Delta T \over T})_{SZ} =
({\Delta T \over T})_{therm} + ({\Delta T
\over T})_{kin}$.

There is considerable interest in the detection of this effect
also because of its potential
in determining the distance of clusters, their peculiar
velocities and in studying the intracluster medium $^{[1-3]}$.
Most of the observations carried out so far (e.g. ${[5]}$) were taken
in the Rayleigh - Jeans (R-J) part of the spectrum,
where the scattering leads to an intensity decrement. 
The more recent radio observations agree in finding this decrement at
centimeter wavelengths
towards A2218, A665, 0016+16, A773, A401, A478, A2142, A2256 $^{[6-9]}$
and Coma $^{[10]}$ and at 2.2{\it mm} towards A2163 
(${[11]}$ and Lange, these proceedings).

Measurements near the planckian peak and on the Wien side have several
advantages: (a) the intensity enhancement relative
to the planckian value is larger than the magnitude of the R-J decrement;
(b) the simultaneous detection of the enhancement (positive) and decrement
(negative) on the same cluster provides an unambiguous signature of
their presence and minimize systematic errors and spurious signals;
(c) sources in the cluster are expected to give a negligible contribution at
high frequency, while radio observations are plagued by the
possible  radio emission from sources within the clusters; (d)
because of the large bandwidth the sensitivity
of bolometer systems is excellent.

\section{The Instrument}

A double channel photometer was built and devoted to the search for the
enhancement and decrement of the S-Z effect. The system works,
in fact, simultaneously
at 1.2 {\it mm} and 2 {\it mm} using two bolometers cooled at 0.3 K
by means of a $^3$He refrigerator. The 2{\it mm} band includes the peak
brightness of the decrement in the
S-Z thermal effect, while the 1.2{\it mm} bandwidth is a compromise between
the maximum value of the enhancement in the S-Z and the atmospheric
transmission. 
The collecting optics, cooled at 0.3 K, define a field of view in the sky
of 44$^{\prime \prime}$ at both frequencies. The beam separation
in the sky was limited by the antenna chopping system and was set to the maximum
chopping amplitude: 135$^{\prime \prime}$.

\noindent
This photometer was built to feed the O.A.S.I. (Osservatorio Antartico
Submillimetrico Infrarosso) telescope installed at the Italian base in
Antarctica $^{[12]}$ and was adapted to the focus of the SEST
and its performance was tested during an observing run
in September 1994.
Details of the instrument can be found in ${[13]}$.

Responsivities, beam shapes and widths were measured with planets and
the main figures measured at the focus are listed in table 1.
Sensitivities are also given in terms of relative change of the
thermodynamic temperatures in one second integration time.

\bigskip
\begin{center}
{{\bf Table 1.} Performances of the photometer at focus}
\end{center}
\begin{center}
\begin{tabular}{cccccccccccc}
\ \\
\hline
\hline
\ \\
&& &$\lambda _c$ & $\Delta \lambda$ & FWHM
 & \multicolumn{1}{c} {noise}
 & \multicolumn{1}{c} {Responsivities}
 & \multicolumn{1}{c} {N.E.T.}
 & \multicolumn{1}{c} {$({\Delta T\over T})_{therm}$} & &\\
&& & ($\mu m$) & ($\mu m$) & ($^\prime$) &
\multicolumn{1}{c}{(nV/$\sqrt{Hz}$)}
 & \multicolumn{1}{c}{($\mu V/K$)}
& \multicolumn{1}{c}{(mK/s)} & \multicolumn{1}{c}{(1s)} & &\\
\ \\
\hline
\hline
\ \\
&&&1200 & 360 & 44 & 45 & 3.0 & 7.5 & 0.010 &&\\
&&&2000 & 580 & 46 & 31 & 1.4 & 10.7 & 0.007 &&\\
\\
\hline
\hline
\end{tabular}
\ \\
\end{center}

\section{Observations}

The targets were selected because of their high
X-ray luminosity in the ROSAT band (0.5$\div$2.4 keV)
$\sim 2- 5 ~ 10^{45}$ erg/s and because of their high redshift
(0.3-0.42). The apparent size
of the core radius, being small, well matches the beam size:
for a core radius of 250$\div$400 kpc in standard cosmologies the apparent
size is (40$\div$60)$^{\prime \prime}$. However, a
chop throw of 135$^{\prime \prime}$ means that at the reference beam
position the ratio between the electron density,
$n_e (\theta=135^{\prime \prime }$),
to its central value, $n_e(\theta=0)$, is 0.25 - 0.35, i.e.
20 - 30 \% of the signal could be lost because of the limited chop throw.

The present paper deals with observations of A2744 and
RXJ0658-5557 carried out in four
different nights during september 1-5 1995.
A total integration time of 10800s and 12400s were spent on A2744 and
RXJ0658-5557 respectively,
and the same integration time was spent on a blank sky
located 15m ahead in right ascension with respect to the source position.
During the observations the sky opacity was very low 
($\tau_{1mm} < 0.1$ with an average value of
$<\tau_{1mm}> = 0.07$, $\tau _{2mm} < 0.05$
with an average value of $<\tau _{2mm}> = 0.03$) and the sky emission
very stable thus producing a very low sky-noise.
%The {\it effective} integration times were 9600s for A2744 and 11000s for
%RXJ0658-5557.

\noindent
In order to get rid of the major sources of noise in this kind of experiment,
fluctuations in the atmospheric emission and systematics from the
antenna, the observing strategy makes use of two combined procedures:
the common three-beam technique, beam-switching + nodding, which gets
rid of the linear spatial and temporal variations in the atmospheric
emission, and the observations of blank sky regions located at right
ascension position of 15m ahead with respect to the location of
the source. This latter implies that for each 10m integration ON the source
(10m integration + overheads give a total tracking time of 15m) a similar
integration is performed on blank sky. The comparison between the two signals
is a measurement of the systematics introduced by the antenna. In fact,
the instrument tracks
the same sky position twice with respect to the local environment,
once ON the source and the other on the blank sky (position OFF).
The choice of 600s of integration ON and OFF the source is a compromise
between the minimization of the time wasted on overheads and the need of
minimizing the atmospheric variations between one observation and the other.

The data analysis procedure is described in detail elsewhere$^{[14]}$.
Briefly, data on each sky position (on the source and on the blank sky)
consist of differential measurements of the antenna temperature,
$\Delta T_{ant}$, with integration chunks lasting 3 x 200s. 200s data
are averaged and a variance is assigned for each of this {\it "scan"}.
%and here it is only briefly outlined.
%Data consist of 200s integration chunks (hereafter called {\it one
%scan}) each containing 20 {\it 10s subscans} taken at two different
%antenna positions, A and B, having the reference beam on the left and on
%the right respectively. The integration sequence
%is ABBAABBAABB...A.
%Spikes due to equipment malfunction and offsets introduced by the electronics
%and the atmosphere were removed with a 2$^{nd}$ order
%polynomial fit and a high-pass Savitzky-Golay filter.
%For each subscan a mean value for the differential antenna temperature,
%$T_A$ or $T_B$, is found by averaging over the 10s.
%The variance, $\sigma^2_{A,B}$,
%is estimated with a procedure of bootstrap resampling.
%This method is widely used to take into account correlation among data
%(induced in this case by the filtering).
%The signal is then obtained by subtracting each couple of subscans
% $\Delta T_i = {{T_A - T_B }
%\over {2}}$ with variance $\sigma ^2 _i = {{ \sigma^2_A + \sigma^2 _B}
%\over {2}}$.

\noindent
%Weighted means are computed for each 600s integration (3 scans)
%on each sky position (when the antenna tracked the source,
%hereafter called $\Delta T_{ON}$, and when the antenna tracked the
%blank sky, hereafter called $ \Delta T_{OFF} $).
Weighted means are computed over the 600s integration 
when the antenna tracked the source (hereafter called $\Delta T_{ON}$)
and when the antenna tracked the blank sky (hereafter called
$ \Delta T_{OFF} $).
Cluster signals are then estimated from the subtraction:
$ \Delta T_{SZ} = \Delta T_{ON} - \Delta T_{OFF} $
and the quadratic sum of the two standard deviations are
used to estimate errors:
$ \sigma _{SZ} ^2 = \sigma ^2 _{ON} + \sigma ^2 _{OFF} $.
Figure 1 ({\it a,b,c,d}) shows the values ON (filled squares)
and OFF (empty squares) of the antenna temperatures
of the two clusters and for both channels.
The solid lines represent the maximum likelihood estimates of the 
$ \Delta T_{SZ} $, while
the dotted lines correspond to the $70\%$ confidence range. 
This was found by computing the 
values of the signal where the likelihood drops
by a factor of $1.71$ from its maximum.
From the figure one can infer that a 3$\sigma$ detection can be claimed
in both channels for RXJ0658-5557: $\Delta T_{1mm} = 0.3$ mK,
$\Delta T_{2mm} = -0.46$ mK, while a signal of $\Delta T_{1mm} = 0.18$ mK, 
is present only at 1{\it mm} for A2744.

\begin{figure}[t]
\epsfxsize=14.cm
\epsfysize=14.5cm
\epsffile{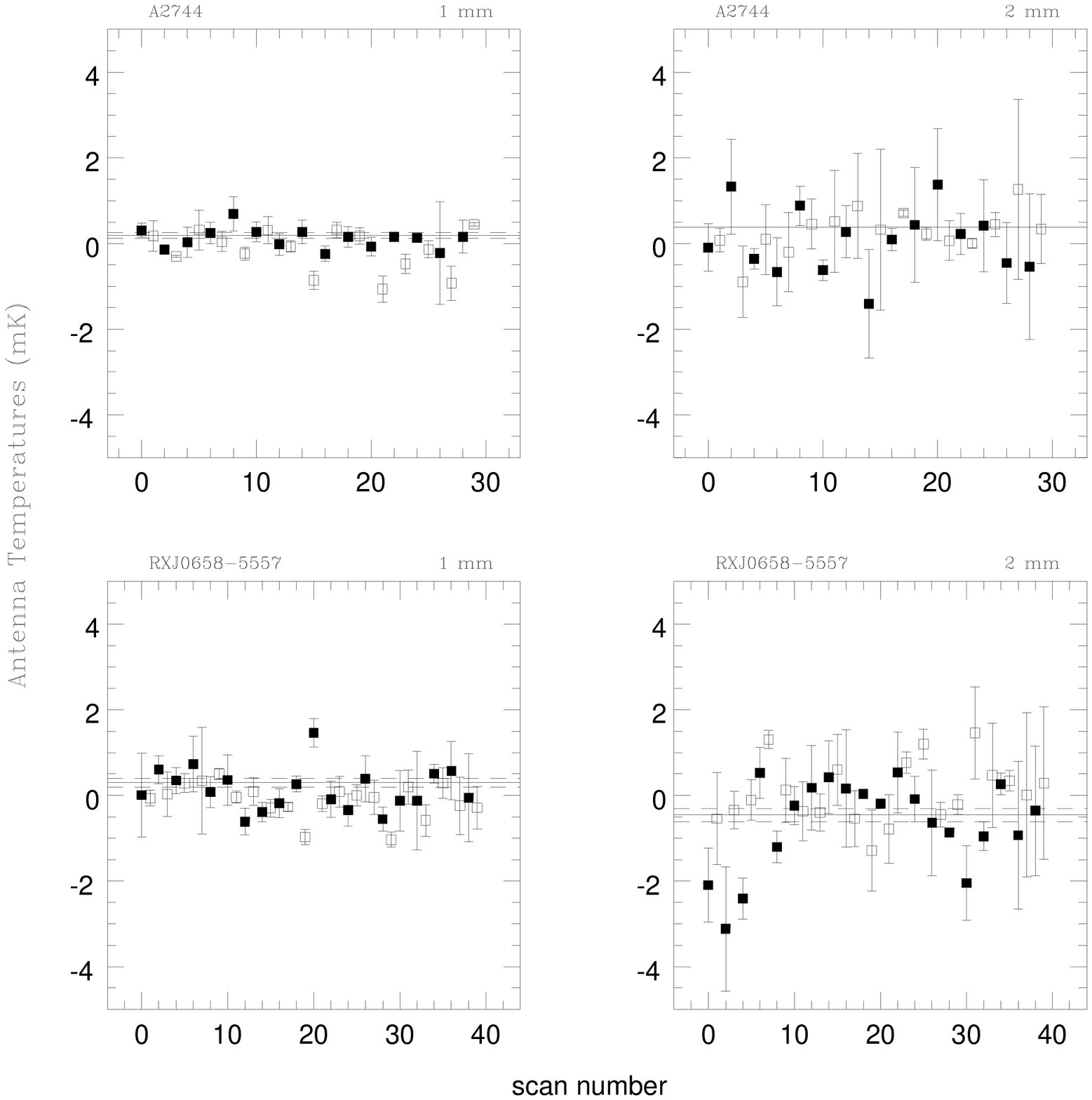}
\caption{Differential antenna temperatures at 1 and 2 {\it mm} for A2744 and RXJ0658-5557.
Filled squares refer to data ON the source, while empty squares to blank
sky regions located 15m ahead in R.A. with respect to the source position.
The cluster signals are estimated by subtracting
$ \Delta T_{SZ} = \Delta T_{ON} - \Delta T_{OFF} $. The maximum likelihood
values of these differences is shown as a solid line while the dashed lines
correspond to 70\% confidence interval. For the 2 {\it mm} signal of A2744
only the 3$\sigma$ upper limit is reported.}
\label{fig:signals}
\end{figure}

To check whether these detections were spurious additional measurements
were carried out to test the behaviour of (a)
the photometer + the electronic chain and (b) the antenna systematics.
The {\it same} observing strategy (on the source and on the blank sky)
was applied in two cases: (a) blocking the photometer entrance window and (b)
following an empty sky region for integration times comparable with those
used for the observations of the sources.
%The behaviour of the former was followed
%blocking the entrance window with an aluminium sheet. The sheet was positioned
%in order to cover the entire window and to
%prevent diffracted radiation to enter the photometer.
%The same observing procedures ON and OFF the source, i,e, the same
%alto-azimuthal paths, have been tracked with this configuration for 10200s.
%No signals are detected with the window covered by the metal sheet and
%there is no trend in the signal as a function of the alto-azimuthal position.
%We conclude that there are no systematics coming from the receiver system.

\noindent
%Antenna systematics were checked by observing a blank sky position with the
%same ON - OFF procedure. Again, the same
%alto-azimuthal paths were tracked for ......s
%Signals were reduced as described above and the resulting averages turn out to
%be: .........

No signals are detected at a level of 0.1 mK (3$\sigma$) so that
we believe that the adopted observing strategy gets rid of
most of the unknown systematics (however one must note that if during the
observation of the sources the antenna, because of a loss of synchronization,
did not track {\it precisely} the same paths relative to the 
local environment, when looking at the source and at the blank sky some,
very small spurious signals can survive).

\section{Discussion}

If the detected signals are due to clusters several questions are
raised. We briefly discuss here only the case of RXJ0658-5557.

Let us assume that the decrement seen at 2{\it mm} is due to the SZ
effect. From this value the {\it expected} 1{\it mm} signal for the thermal effect
can be easily computed by convolving eq.1 with the beam shape, the optics
transmission and the atmospheric spectrum. However, the difference with the
observed value is larger than a factor of 2 and this can be explained
either with a large peculiar velocity of the cluster, of the order
of -1000 km/s, or with a source contaminating mainly
the 1{\it mm} channel. To our knowledge there is no determination of the
peculiar velocity from optical data and we
are not able to verify the first hypothesis. The latter can be checked
in several ways.

(1) No sources in the IRAS Faint Source Catalogue
or in the 5GHz NRAO survey are present at the position of the main beam
and/or of the reference beams.

(2) If we scale the 60$\mu m$ IRAS flux limit of 240 mJy at 1.2{\it mm} by using
the average flux of nearby spirals $^{15}$, we find that a normal
spiral would give rise to a signal not larger than 0.02 mK in
antenna temperature.

(3) If we assume a contribution of many unresolved sources fluctuating in
the beam and take the estimation made by Franceschini et al. ([${16}$]), the
expected signal will be not larger than 0.02 mK.

(4) Irregular emission from the Galactic cirrus can also give origin to a
signal at these wavelengths. If we take
the estimation by Gautier et al. ([${17}$]) and extrapolate the 100$\mu m$ flux
at 1.2 {\it mm} using the average Galactic spectrum a maximum signal of 0.02 mK
is found.

(5) If part of the signal is due to CMB anisotropies at these scales,
it will be hard to disentangle them from the S-Z kinematic effect
since this latter has a spectrum identical to that of the anisotropies
(see e.g. [${18}$]).

We conclude that it is very likely that part of the signal at 1.2 {\it mm} is
due to one or more sources but at present it is hard to identify it. This
point however deserves further investigations and it will be the goal of future
research.

\vskip 1cm
The authors are indebted to the ESO/SEST teams at La Silla and in particular
to Peter de Bruin, Peter Sinclair, Nicolas Haddad and Cathy Horellou.
This work has been partially supported by the P.N.R.A. (Programma
Nazionale di Ricerche in Antartide). P.A. warmly thanks ESO for hospitality
during 1995, when part of this work was carried out.

\end{document}